\documentclass[12pt]{article}
\usepackage{amsmath}
\usepackage{amssymb}
\usepackage{graphicx}
\newcommand{\be}{\begin{equation}}
\newcommand{\ee}{\end{equation}}
\newcommand{\ba}{\begin{eqnarray}}
\newcommand{\ea}{\end{eqnarray}}
\newcommand{\bc}{\begin{center}}
\newcommand{\ec}{\end{center}}

\begin{document}

\begin{center}
{

\bf Metal-insulator transition\\ induced by fluctuations of the magnetic potential\\
in semiconductors with magnetic impurities}
 \end{center}
\vspace{1cm}

\centerline{E.Z. Meilikhov\footnote{$^)$ e-mail: meilikhov@imp.kiae.ru}$^)$, R.M. Farzetdinova}
\medskip
\centerline{\small\it Kurchatov Institute, 123182 Moscow, Russia}
  \vspace{1cm}

\begin{center}
\begin{tabular}{p{15cm}} \footnotesize\quad

We investigate the metal-insulator transition occurring in semiconductors with magnetic impurities when lowering
temperature. In contrast to the usually considered percolation transition in the non-uniform medium induced by
the localization of charge carriers in the fluctuating \emph{electric }potential, the studied transition is
connected with their localization in the fluctuating \emph{magnetic} potential produced by magnetized impurities
(more accurately -- in the combined fluctuating potential). When decreasing temperature, the magnetization of
magnetic impurities in the semiconductor becomes higher and even at the invariable (temperature independent)
amplitude of the electric potential, the magnetic component of the total potential increases. With increasing
fluctuation amplitude, the Fermi level of charge carriers sinks deeper and deeper into the growing tail of
density of states  until it falls under the percolation level. For that, fluctuations of the total potential
have to run up to some critical value. On reaching that value, the transition occurs from the metal conductivity
to the activation one (the metal-insulator transition).

\end{tabular}
\end{center}
\vspace{1cm}

\centerline{\bf Introduction}
\medskip

The role of large-scale fluctuations of the electric potential in traditional (\emph{non-magnetic}) doped
semiconductors is well known~\cite{1}. Such a fluctuating potential appears usually in highly-compensated
semiconductors where concentrations of charged impurities (donors and acceptors) are high and the concentration
of screening mobile charge carriers is low, that results in a large screening length $\ell_s$, defining the
spatial scale of electric potential fluctuations.  In that case, the average amplitude of the fluctuation
potential is also high that leads to the localization of charge carriers and results in the activation character
of the system conductivity: it is controlled by the thermal activation of charge carriers from the Fermi level
to the percolation level and falls down exponentially with  lowering temperature.  In the absence of the
impurity compensation, the charge carrier concentration is so high that any perturbations of the electrostatic
nature are effectively screened, and  the spatial scale of the potential coincides with the extent of impurity
density fluctuations. The depth of such a short-scale potential relief is relatively shallow and does not lead
to the charge carrier localization -- the conductivity  keeps to be metal one.

In diluted (but nevertheless, highly-doped) \emph{magnetic} semiconductors (of Ga$_{1-x}$Mn$_x$As type), in
addition to above mentioned fluctuations of the electric potential, the new perturbation source appears --
specifically, fluctuations of the  "magnetic potential" concerned with fluctuations of the local magnetization
in such a semiconductor~\cite{2}. That potential is, in fact, the potential of the exchange interaction of
mobile charge carriers with magnetic impurities~\cite{3} (for instance, via the RRKKY mechanism) which
fluctuates in accordance with fluctuations of the concentration and the local magnetization of those impurities.

Within  the "wells" of the magnetic potential, mobile charge carriers with a certain spin direction are
accumulated while the carriers of the opposite spin direction are pushed out. The spatial scale $\ell$ of
magnetic fluctuations is now determined not by the electrostatic screening but by the characteristic length of
the magnetic interaction of impurities and the correlation length of their arrangement in the semiconductor
bulk. However, in diluted magnetic semiconductors, there is usually  $\ell\sim\ell_s$ and, thus, spatial scales
of the magnetic (exchange) and Qoulomb potentials agree closely.

That means the constructive superposition of both reliefs, and so the average total amplitude of the potential
relief becomes to be higher. The medium arises where the concentration and the spin polarization of charge
carriers are strongly non-uniform, and the degree of that non-uniformity is substantially defined by the local
magnetization of the system.

Increasing magnetization with lowering temperature promotes strengthening the spatial localization of charge
carriers and in a number of cases could stimulate the metal-insulator transition~\cite{4}. Percolative metal
conductivity, characteristic for non-uniform systems, changes into the conductivity of the activation type. That
occurs when under some external factors (such as temperature, magnetic field, etc.) the Fermi level falls below
the percolation level. One of possible mechanisms is as follows. The  fluctuating potential leads to appearing
the  density of states tail into which both the percolation and Fermi levels are pulled. However, rates of those
levels' movement are different, and if they change the relative position the metal-insulator transition occurs.
The possibility of such a model is investigated in the present work.

Similar transitions have been repeatedly observed in various systems with magnetic impurities coming into
ferromagnetic state at lowering temperature. Thus, for example, in~\cite{5} the resistivity temperature
dependencies of the magnetic semiconductor Cd$_{0.95}$Mn$_{0.05}$Se (in which the electron concentration $n$
being varied by means of the additional doping by In), have been studied. At relatively low electron
concentrations  ($n\lesssim10^{18}$~cm$^{-3}$),  the resistivity has sharply increased at lowering temperature,
that could be interpreted as the transition in the insulator state. The lower $n$, the higher the specific
temperature of that transition. Analogous effect has been observed for the compound  Ga$_{1-x}$Mn$_x$As with
$x\sim0.02$~\cite{6}. Similar process occurs also in Ge, strongly doped  (by using ion implantation) with
Mn-atoms, whose relative concentration being of 2-4\%~\cite{7}.

We believe that in all those cases the nature of the metal-insulator transition is the same, namely, the
localization of charge carriers in the fluctuating magnetic potential which amplitude increases with lowering
temperature along with  the magnetization of magnetic impurities. Investigating and describing the
mechanism of that transition is the object of the present paper.   \\
\bigskip

\centerline{\bf Fluctuations of the magnetic potential}
\medskip

It is convenient to characterize the non-uniform magnetization $M({\bf r})$ of impurities (to be definite, Mn
atoms are borne in mind below) by the coordinate dependent local magnetization  $j({\bf r})\equiv M({\bf
r})/M_s$, where $M_s$ is the saturation magnetization. Non-zero local magnetization ($0\leq j\leq 1$ ) of Mn
atoms with the spin $S_{\rm Mn}=5/2$ leads to the non-uniform local spin polarization of holes which is
reflected in the fact that the local concentration $p^-({\bf r})$ of holes with the spin being antiparallel to
the local impurity magnetization exceeds the concentration $p^+({\bf r})$ of holes with the opposite spin
orientation. At that, the local  degree of the hole polarization $\xi({\bf r})=[p^-({\bf r})-p^+({\bf
r})]/p({\bf r})$ is non-zero (\,$p({\bf r})=p^-({\bf r})+p^+({\bf r})$ is the total local hole concentration.)
The hole polarization could be interrelated formally with the action of the effective spin-dependent magnetic
potential~\cite{2}
 \be\label{1}
U_{\rm mag}({\bf r})=xN_0a^3J_{pd}\,\sigma S_{\rm Mn}j({\bf r}),
 \ee
where $N_0\approx8\cdot10^{21}$ cm$^{-3}$ is the concentration of Ga-sites in the GaAs-lattice, $x$ is the
fraction of those sites being taken by Mn atoms, $a$ is the lattice constant, $\sigma=\pm 1/2$ is the hole spin,
$J_{pd}=1.2$ eV is the energy of the exchange interaction between mobile holes and localized $d$-electrons of Mn
atoms~\cite{6}. In accordance to (\ref{1}), major holes (possessing the preferred spin orientation) are
accumulated in the high magnetization areas, whereas minor ones, conversely, are ejected in low magnetization
areas.

One could found the relation between amplitudes of the magnetic and electrostatic potentials for the impurity
concentration fluctuation of the dimension $R_f$ by noticing that the latter could be estimated as  $U_{\rm
el}\sim (e^2/\kappa R_f)(4\pi xN_0 R_f^3)^{1/2}\sim e^2/\kappa R_f$ ($\kappa\sim10$ is the dielectric
susceptibility, $\ell_s\sim a$)~\cite{1}. Then
$$\frac{U_{\rm mag}}{U_{\rm el}}\sim
\frac{jJ_{pd}}{(e^2/\kappa R_f)}\sim \left(\frac{R_f}{a}\right)j,
$$
so at $j\sim 1$  we have $U_{\rm mag}\gg U_{\rm el}$ for fluctuations of the dimension $R_f\gg a$. In that case,
it is unacceptable to ignore the magnetic potential. However, more important characteristics of the "magnetic
relief" is  the average amplitude $\langle\Delta U_{\rm mag}\rangle$ of the potential fluctuations (cf. below).

Let, in the absence of the magnetization ($j=0$), the Fermi energy of charge carriers (holes) be
$\varepsilon_F$. The magnetic potential (\ref{1}) originating with appearing magnetization results in the
splitting of the hole band into two spin sub-bands with effective Fermi energies (reckoned from edges of those
sub-bands) $\varepsilon_F+U_{\rm mag}({\bf r})$ and $\varepsilon_F-U_{\rm mag}({\bf r})$. Two different
effective Fermi momenta
 \be\label{2}
k_F^{\pm}({\bf r})=k_F\left(1\pm\frac{U_{\rm mag}({\bf r)}}{\varepsilon_F}\right)^{1/2},
 \ee
correspond to those energies, where $k_F=(2m^*\varepsilon_F/\hbar^2)^{1/2}$, $m^*\approx0.5m_0$ is the effective
hole mass (at $U_{\rm mag}({\bf r})/\varepsilon_F>1$, $k_F^-({\bf r})=0$).

Under the conditions of the uniform magnetization ($j={\rm Const}$) the momenta  $k_F^\pm $ are independent of
coordinates and, taking into account the spin splitting of the hole band, the expression for the interaction
energy $w(\rho)$ of  two magnetic atoms, spaced by the distance $\rho$, could be written  as
 \be\label{3}
w(\rho)=I_0 \Phi(\rho),
 \ee
where the range function  $\Phi(\rho)$ depends on the mechanism of the indirect interaction of magnetic
impurities and, by the example of the RKKY interaction, reads~\cite{8}
 \be\label{4}
 \Phi(\rho)=\left(\frac{a}{\rho}\right)^4
[\sin\theta^+(\rho)-\theta^+(\rho)\cos\theta^+(\rho)+\sin\theta^-(\rho)-
\theta^-(\rho)\cos\theta^-(\rho)]\exp(-\rho/\ell),
 \ee
 \be\label{5}
I_0=\frac{1}{(4\pi)^3}\left(\frac{ma^2}{\hbar^2}J_{pd}^2\right),\quad \theta^\pm(\rho)=2k_F^\pm \rho.
 \ee
The exponential factor in (\ref{4}) allows for the finite length $\ell$ of the hole spin relaxation~\cite{8}
(being equal to their collision length, in the simplest case).

Notice, with a view of the present work the concrete form of the indirect interaction of magnetic impurities is
not a matter of principle. RKKY interaction is  used as a model only, permitting to carry later calculations up
"to digits".

Generalization of the function (\ref{4}) for the case of the non-uniform magnetization is made by replacing
phases $\theta^\pm(\rho)$ with their average magnitudes
 \be\label{6}
\theta^*_\pm(\rho)=2\int\limits_0^\rho k_F^\pm
(s)ds=2k_F\int\limits_0^\rho \sqrt{1\pm Aj(s)}ds,
 \ee
where $A=xN_0a^3J_{pd}\sigma S_{\rm Mn}/\varepsilon_F$, and the integration is executed along the line
connecting impurities~\cite{8}. Then the function $\Phi(\rho)$ transforms into the functional of the spatially
non-uniform magnetization:
 \ba\label{7}
\Phi(\rho)\to\hat F [Aj({\bf r}), \rho]
\equiv\hspace{90pt}\nonumber\\
\label{7}\equiv\left(\frac{a}{\rho}\right)^4 [\sin\theta_+^*
(\rho)-\theta_+^*(\rho)\cos\theta_+^*(\rho)+\sin\theta_-^*(\rho)-\theta_-^*(\rho)\cos\theta_-^*(\rho)]e^{-\rho/\ell}.
 \ea
At low magnetization ($j\ll 1$), the magnetic potential (\ref{1}) is also small and the relation (\ref{7}) takes
the standard form.

As could be seen from Eq. (\ref{6}), the  non-uniformity of the magnetization plays the significant role in that
case only when $Aj\sim1$ (i.e., the magnetic potential is comparable with the Fermi energy). Let us estimate the
parameter $A$ for  Ga$_{1-x}$Mn$_x$As semiconductor. Due to the compensation, the hole concentration $p$ is
always lower than the concentration  $xN_0$  of  Mn atoms (acceptors) positioned in Ga-sites. Nevertheless,
$p/(xN_0)\gtrsim 0.3$ at $x=0.05$. The estimation for that case gives $A\sim 1$~\cite{8}. Thus, the
magnetization non-uniformity should be taken into account in every system area with the local magnetization
$j\sim 1$.

Further calculations relate to the simplest case when the magnetization has the same direction anywhere (for
instance, due to the strong uniaxial magnetic anisotropy), i.e., $j({\bf r})$ is the scalar value.

The energy $W_{\rm RKKY}$ of the indirect interaction of a given spin $S_i$ with its surroundings is determined
by the sum $W_{\rm RKKY}=\sum_{j=1}^\infty w(\rho_{ij})$. The distance $\rho_{ij}$ could not be smaller that the
distance  $a_0=N_0^{-1/3}$ between two neighbor sites of Ga sublattice accessible for magnetic impurities (for
the diluted Ga$_{1-x}$Mn$_x$As  semiconductor $a_0=a/\sqrt{2}$, where $a\approx5$\AA\, being the lattice
constant). In the continual approximation the sum could be replaced by the integral
 \be\label{8}
W_{\rm RKKY}(\textbf{r})=I_0N_0\int \hat F [Aj({\bf r}'), |{\bf
r}-{\bf r'}|]\,x(\textbf{r}')j(\textbf{r}')d^3\textbf{r}',
 \ee
where the integration executed over the volume occupied by impurities.

To obtain qualitative results, below  we will not go beyond the simple model range function  $\Phi(\rho)$,
depending on the distance between impurity atoms only and coinciding with that for the uniform system. Then, in
spherical coordinates, where a given impurity atom is situated at the distance $h$ from the coordinate origin,
 \be\label{9}
W(h)=I_0N_0\int\limits_r\int\limits_\varphi\int\limits_\theta
\Phi[\rho(h,r,\varphi,\theta)]x(r,\varphi,\theta)j(r,\varphi,\theta)r^2\sin\theta
dr d\varphi d\theta,
 \ee
where $\rho(h,r,\varphi,\theta)=(r^2+h^2-2rh\sin\theta\cos\varphi)^{1/2}$.

Hereinafter, we take an interest in spatial dependencies of the impurity concentration $x(r)$ and magnetization
$j(r)$ only. Taking into account the angle dependencies of those values could not lead to the principal
variation of physical parameters of the length dimension, such as  "screening length" of the point magnetic
perturbation (delta-like burst of the magnetic impurity concentration). Therefore, in Eq.~(\ref{9}) we keep the
spatial dependence of above mentioned parameters only (or, in other words, replace them by values, averaged over
angles). Then Eq.~(\ref{9}) takes the form
 \be\label{10}
W(h)=I_0N_0\int\limits_r\int\limits_\varphi\int\limits_\theta
\Phi[\rho(h,r,\varphi,\theta)]x(r)j(r)r^2\sin\theta dr d\varphi d\theta,
 \ee
where the integration range is defined by the condition $\rho(h,r,\varphi,\theta)\ge a_{\rm min}$. For the
function $\Phi(\rho)$ which is non-divergent at $\rho\to0$,  integrating could be performed over the whole
space, and the convergence of the integral (\ref{10}) is guaranteed by the fast decay of the range function.

In the latter case, the self-consistent equation $j(h)=B_S[W(h)/kT]$, defining the local magnetization ($B_S$ is
the Brillouin function for the spin $S$), reads as follows
 \be\label{11}
j(h)=B_S\!\left[\frac{1}{\tau}\left(\frac{1}{a_0^3}\int\limits_{r=0}^\infty
K(h,r)\,x(r)j(r)\,r^2 dr\right)\right],
 \ee
where \be\label{12}
 K(h,r)=\int\limits_{\varphi=0}^{2\pi}\,\int\limits_{\theta=0}^\pi
\mathfrak{h}(\rho/a_0)\Phi[\rho(h,r,\varphi,\theta)]\sin\theta\,d\varphi\,
d\theta,
 \ee
$\tau=kT/I_0$ is the reduced temperature, $\mathfrak{h}(t)$ is the unit Heaviside function. That is  the
integral equation defining the spatial dependence of the local magnetization $j(r)$ at a given spatial
distribution $x(r)$ of the magnetic impurity concentration. For the uniform doping, when $x(r)=x_0$, the system
magnetization is also uniform: $j(r)=j_0$.

For the average impurity concentration $x_0$, the number of impurity atoms in the volume $V$ is, in average,
$\bar N= x_0N_0 V$ with the standard deviation $\bar N^{1/2}$ from that value. The relevant fluctuation of the
relative concentration equals $\Delta x=\pm\sqrt{x_0(a_0^3/V)}$, or
 \be\label{13}
 \Delta x=\pm(a_0/R_f)^{3/2}\sqrt{(3/4\pi)x_0}
 \ee
for the spherical fluctuation of the radius $R_f$.

Let us consider the magnetic perturbation in the uniform semiconductor as the Gauss spherically symmetrical
fluctuation of the magnetic impurity concentration in the coordinate origin: $x(r)=x_0+\delta x(r)$, where
$\delta(x)=\Delta x\exp(-r^2/R_f^2)$. The "response" of the system to that perturbation appears as the
non-uniformity of its magnetization $j(r)$ about a point of that fluctuation. The size of the relevant
non-uniform area and the magnitude of the magnetization deviation from the bulk value  $j_0$ could be found with
the help of Eq.~(\ref{11}).

That phenomenon is the magnetic analog of screening electric charges  in strongly doped semiconductors with
impurity fluctuations. In this case, the average amplitude and the spatial scale of the fluctuating
\emph{electric} potential is determined by optimal impurity fluctuations with the size close to the screening
length~\cite{1}. The same characteristics of the fluctuating \emph{magnetic} potential are defined by the
characteristic length of the magnetic impurity interaction (cf.  (\ref{3}), (\ref{4})) and to a large extent, by
temperature, as well (cf. (\ref{11})).

Fig.~1 demonstrates spatial perturbations $U_{\rm mag}(h)$ of the magnetic potential, generating by  spherical
fluctuations of the impurity concentration (situated in the coordinate origin) whose amplitude and size are
interrelated by Eq.~(\ref{13}). They are calculated in the course of the numerical solution of the equation
(\ref{11}) by successive approximations' method (for some realistic set of parameters  $x_0$, $k_Fa^3$,
$\ell/a$). At the large distance from the origin, the magnetic potential tends to its bulk value characteristic
for the uniform medium with the impurity concentration $x_0$. The most deviation from that value is observed,
naturally, near the origin where the center of the impurity fluctuation is located.  The variation of the
magnetic potential
 \be\label{DU}
  \Delta U_{\rm mag}=U_{\rm mag}(h=0, \Delta x>0)-U_{\rm mag}(h=0, \Delta x<0),
 \ee
defined as the difference between magnetic potentials in centers of the positive ($\Delta x>0$) and negative
($\Delta x<0$) fluctuations, depends, naturally, on the amplitude $\Delta x$ (or on the size $R_f$, cf.
(\ref{13})) of the fluctuation. It is hereinafter important that the average  amplitude of the magnetic
potential fluctuations coincides with $\Delta U_{\rm mag}$ on the order of value and defined by the
characteristic spatial scale of impurity concentration fluctuations.

In~\cite{9} that scale is identified with the correlation length of impurity arrangement which (taking into
account the mutual attraction of Mn atoms in GaAs) is estimated as $R_f\approx (3-5)a_0$. Basing upon that
estimate, we accept the typical fluctuations being of the radius $R_f=5a_0$. At the average impurity
concentration $x_0=0.05$, that results in the concentration fluctuation equal to $\Delta x\approx 0.03$ (cf.
(\ref{13})). For those fluctuations, according to Fig.~1  $\Delta U_{\rm mag}\approx 0.03J_{pd}\approx 30$~meV.
It is just the average amplitude of magnetic potential fluctuations (its difference between maximum and minimum)
in the considered diluted magnetic semiconductor. Its temperature dependence is shown in Fig.~2 -- magntic
fluctuations arise with appearing non-zero magnetization and with further temperature lowering their amplitude
is saturated in compliance with the magnetization saturation.

The relation between amplitudes of the magnetic and electrostatic potentials for the fluctuation of impurity
density of the radius $R_f$ could be found, noticing that the latter could be estimated as $\gamma_e\sim
(e^2/\kappa R_f)(\Delta x N_0 R_f^3)$~\cite{1}. Herefrom, it follows  $$ \frac{\Delta U_{\rm mag}}{\gamma_e}\sim
\left(\frac{R_f}{a}\right)j,$$ so that $\Delta U_{\rm mag}\gtrsim\gamma_e$ at $j\sim 1$.
\bigskip

\centerline{\bf Metal-insulator transition}
\medskip

If the magnetization is low ($j\ll 1$), the magnetic potential and spin polarization of charge carriers (holes)
could be neglected. In that case, their transport is defined by the possibility of Anderson localization in the
random electrostatic potential and could be of the metal or thermoactivation (insulator) type depending on the
mutual disposition of the Fermi level $\varepsilon_F$  and the percolation one $U_p$. The latter is determined
by the condition $\int_{-\infty}^{U_p}F(U)dU=\theta_p$, where $F(U)$ is the distribution function of random
potential, $\theta_p\approx0.17$ is the fraction of the space where the potential  $U<U_p$~\cite{1}. For the
Gauss function $F(U)=(2\pi\gamma_e^2)^{-1/2}\exp(-U^2/2\gamma_e^2)$  (symmetrical relative to the
level~$U=0$,~corresponding to the edge of the mobile charge carriers' band)  we have $U_p\approx-0.95\gamma_e$.

In the diluted magnetic semiconductor, which is a slightly-compensated doped semiconductor, the amplitude of the
electrostatic potential is relatively small and the charge carrier concentration is so high that the Fermi level
is positioned above the percolation level. That results in the metal conductivity.

Engaging the magnetic fluctuation potential at low temperatures changes the potential relief significantly.
Since at low temperatures current carriers are strongly polarized (those ones prevail, whose spin is
antiparallel to spins of polarized magnetic impurities), it is sufficient to take into account the magnetic
potential for such carriers only (that means $\sigma=-1/2$ in~(\ref{1})~). Exactly such a simple model is
applied below. Statistic properties of the total potential,  which is the sum of the electrostatic and mentioned
magnetic potentials, are described by the distribution function, whose halfwidth enlarges (comparing to the
initial distribution function of the electric potential) approximately by the magnitude $\gamma_m=\Delta U_{\rm
mag}/2$:
 \be\label{14}
 F(U)=\frac{1}{\sqrt{2\pi}(\gamma_e+\gamma_m)}\exp
 \left[-\frac{1}{2}\left(\frac{U}{\gamma_e+\gamma_m}\right)^2\right].
 \ee
Then, with increasing the magnetic potential $\Delta U_{\rm mag}$, the percolation level (defined by the former
condition  $\theta_p=0.17$) drifts downward according to the simple linear low
 \be\label{15}
 U_p\approx-0.95\gamma,\, \gamma=\gamma_e+\gamma_m.
 \ee

The position of the Fermi level depends also on the total amplitude $\gamma$ of the fluctuation potential:
appearing tail of the density of states "pulls" it down. Unlike the density of states
$g_0(\varepsilon)=(2m^*)^{3/2}\sqrt{\varepsilon}/2\pi^2\hbar^3$ in the uniform medium, turning into zero at
$\varepsilon\le 0$ ($\varepsilon=0$ corresponds to the edge of the charge carrier band), the tailed density of
states $g(\varepsilon)$ in a medium with fluctuating impurity concentration is determined by the
relation~\cite{1}
 \be\label{16}
 g(\varepsilon)=\frac{(2m^*)^{3/2}\sqrt{\gamma}}{2\pi^2\hbar^3}\, G_0(\varepsilon/\gamma),
 \ee
where
 \be\label{17}
G_0(X)=\frac{1}{\sqrt{\pi}}\int\limits_{-\infty}^X e^{-y^2}(X-y)^{1/2}dy.
 \ee
At the invariable concentration $n$ of degenerate charge carriers, the Fermi level $\mu$ shifted due to the
fluctuations (of the average amplitude $\gamma$) could be found from the relationship
$$
n=\int\limits_0^{\mu_0} g_0(\varepsilon)d\varepsilon=\int\limits_{-\infty}^\mu g(\varepsilon)d\varepsilon
$$
($\mu_0$ is the Fermi energy in the uniform medium), which leads to the equation
 \be\label{18}
\frac{1}{\sqrt{\pi}}\int\limits_{-\infty}^{\mu/\gamma}\left[\,\int\limits_{-\infty}^X
e^{-y^2}(X-y)^{1/2}dy\right]dX=\frac{2}{3}\left(\frac{\mu_0}{\gamma}\right)^{3/2}.
 \ee
The result of solving Eq. (\ref{18}) is represented in Fig.~3. At a small fluctuation amplitude, the Fermi
energy  coincides practically with that for the uniform medium ($\mu\approx\mu_0$). However, with increasing
amplitude $\gamma$ it drifts quickly down into the range of developing tail of density of states.

If drifting the Fermi level $\mu$ can result in changing the character of the conductivity (metal or activation)
depends on how  the percolation level  $U_p$\, shifts  (cf. (\ref{15})).  Fig.~4 shows the arrangement of those
two levels as a function of the total fluctuation amplitude $\gamma$. Evidently, at $\gamma\approx10\mu_0$ their
mutual disposition changes: at small $\gamma$ values, the Fermi energy level is above the percolation level and
at large $\gamma$, it lies lower. That means the transition from the metal conductivity to the activation one.

Such a transition could happen under the action of different factors. One could influence either electrostatic
component $\gamma_e$ of the total fluctuation amplitude $\gamma$, or the magnetic component $\gamma_m$. For
example, the well-known metal-insulator transition  observing under varying the compensation degree in strongly
doped compensated semiconductors~\cite{1} relates to the first case.

The second case corresponds to the above considered temperature transition in magnetic semiconductors associated
with the dependence of the component  $\gamma_m$ on the local magnetization varying with temperature (cf.
(\ref{DU})). Qualitative notion concerning the temperature dependence of the resistivity $\rho(T)$ of the
magnetic semiconductor could be obtained with using the simple model relation
 \be\label{19}
\rho(T)=\left\{\begin{tabular}{lll}
               $\rho_0$&\!\!\!\!\!,&$U_p<\mu$\\
               $\rho_0\exp[(U_p-\mu)/kT]$&\!\!\!\!\!,&$U_p>\mu$
               \end{tabular}\right.,
 \ee
if one specifies the average amplitude $\gamma_e$ of\, "seed" electrostatic fluctuations.

Dependencies $\rho(T)$, found with the described procedure, are presented in Fig.~5. They demonstrate the
metal-insulator transition induced by the magnetic potential fluctuations at lowering temperature, and look
qualitatively like experimental dependencies $\rho(T)$ for considered systems (see the insert in Fig.~5; slow
temperature upgrowth of Ge$_{0.98}$Mn$_{0.02}$ resistivity in the metal state is connected with carrier
scattering by phonons). The characteristic temperature of such a transition depends significantly on the
fluctuation amplitude  $\gamma_e$ of the initial \emph{electrostatic} potential: the higher $\gamma_e$, the
lower that temperature. Since $\gamma_e\propto x_0^{1/2}$ (cf (\ref{13})), the transition temperature should
decrease with increasing the impurity concentration $x_0$ that also agrees with experiments~\cite{5,6,7}.
 \bigskip

\centerline{\bf Conclusions}
\medskip

In conclusion, we have considered the metal-insulator transition happening with lowering temperature in
semiconductors with magnetic impurities. Unlike the percolation transition in the non-uniform medium induced by
the charge carriers localization in the fluctuating \emph{electric} potential of charged impurities~\cite{1} (of
the average amplitude~$\gamma_e$), the considered transition is connected with the localization in the
fluctuating \emph{magnetic} potential (of the average amplitude~$\gamma_m$) generated by the impurity
magnetization, or more accurately -- in the total fluctuation potential (of the average amplitude~
$\gamma=\gamma_e+\gamma_m$). Since with lowering temperature the magnetization of magnetic impurities in the
semiconductor increases, even under the invariable (temperature independent) electric potential
amplitude~$\gamma_e$ the magnetic component~$\gamma_m$ of the total potential enlarges. The metal-insulator
transition occurs when potential fluctuations becomes so high that the Fermi level~$\mu$ falls into the range of
localized states below the percolation level $U_p$.

The transition temperature is defined by establishing the condition  $\mu(T)\leqslant U_p(T)$, where, according
to~(\ref{15}), $U_p(T)=0.95\gamma(T)$, and $\mu$ is determined by the equation~(\ref{18}). Its solution could be
written in the form $\mu(T)=F[\gamma(T)]$, where $F(\gamma)$ is some function whose plot (for a certain set of
parameters) is represented in Fig.~4. Thus, the transition temperature $T_c$ is found from the condition
$F[\gamma(T)]< 0.95\gamma(T)$, or $\gamma (T)>\gamma_c$, where $\gamma_c$ is some critical magnitude of the
average amplitude of total potential fluctuations. The reason for the transition consists in that with
increasing fluctuation amplitude the Fermi level sinks deeper and deeper into the growing tail of the density of
states until (at the beginning of the condition $\gamma>\gamma_c$) it falls below the percolation level. At
that, the transition from the metal conductivity to the thermoactivation one happens (the metal-insulator
transition).
\bigskip

This work has been supported by Grants \#\# 09-02-00579, 09-02-92675
of the Russian Foundation of Basic Researches.

\newpage
\newpage\centerline{\bf Figure captions}
\bigskip

Fig. 1. Spatial perturbations $U_{\rm mag}(h)$ of the magnetic potential generated by  spherical fluctuations of
the impurity concentration (situated in the coordinate origin). Upper curves are for "positive" fluctuations
($\delta x>0$), lower ones -- for "negative" fluctuations ($\delta x<0$). Accepted parameters: $x_0=
0.05$, $\tau = 0.05$, $k_Fa = 0.1$, $\ell = 5a$.\\

Fig. 2. Temperature dependence of the average amplitude $\Delta U_{\rm mag}$ for  magnetic potential
fluctuations connected with spherical impurity density fluctuations of the radius  $R_f=5a_0$.
Accepted parameters: $x_0=0.05$, $\tau = 0.05$, $k_Fa = 0.1$, $\ell = 5a$.\\

Fig. 3. Shifting the Fermi energy $\mu$ as a function of the total potential fluctuation amplitude $\gamma$.
Dashed line is the unshifted Fermi energy $\mu_0$ (in the absence of fluctuations). \\

Fig. 4.  Mutual arrangement of the Fermi level $\mu$ and the percolation level $U_p$ as a function of the total
potential fluctuation amplitude $\gamma$. Dashed line separates areas of the metal and activation (\emph{I})
conductivities.\\

Fig. 5. Model temperature dependencies of the magnetic semiconductor resistivity $\rho(T)$  for two magnitudes
of the electric potential fluctuation amplitude $\gamma_e$. Accepted parameters: $x_0=0.05$, $k_Fa = 0.1$, $\ell
= 5a$, $J_{pd}=1.2$ eV, $\mu_0=5$ meV. In the insert: experimental temperature dependencies of the resistivity
for Ga$_{0.985}$Mn$_{0.015}$As~\cite{6} and Ge$_{0.98}$Mn$_{0.02}$~\cite{7}.

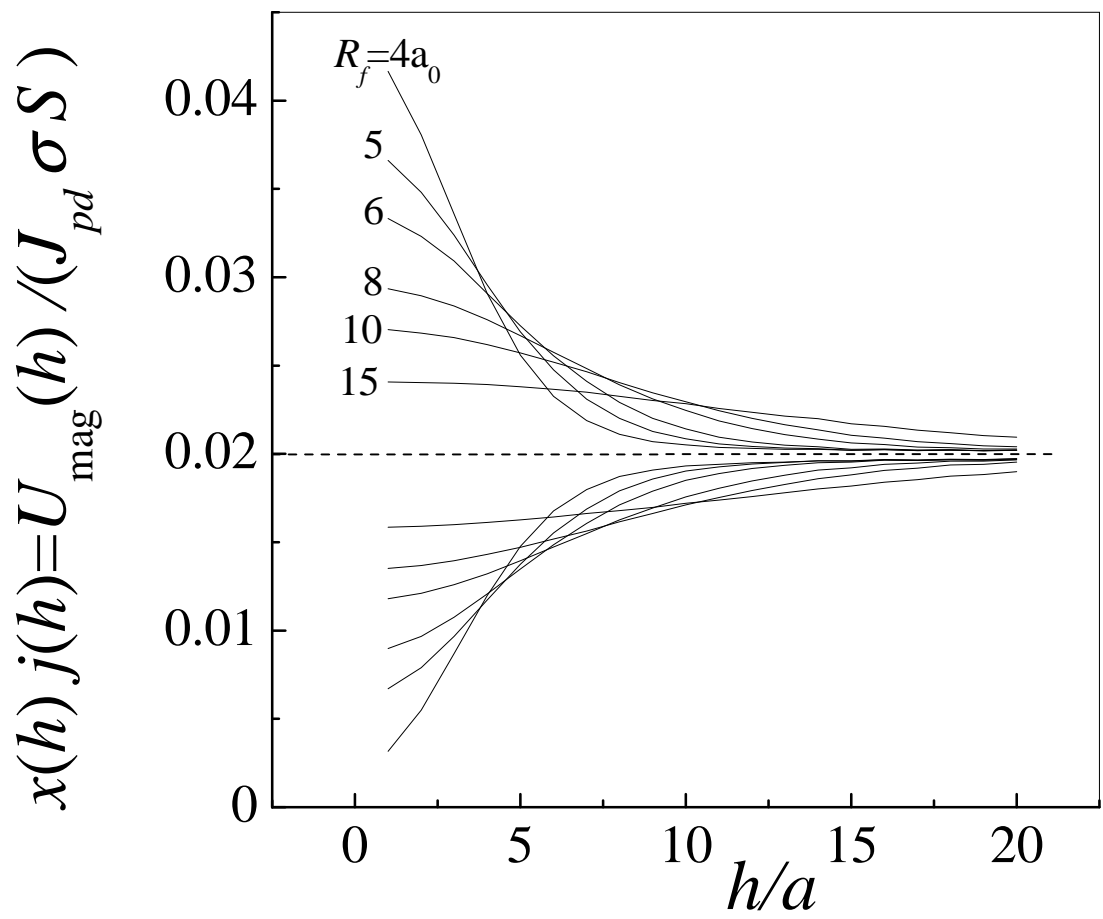
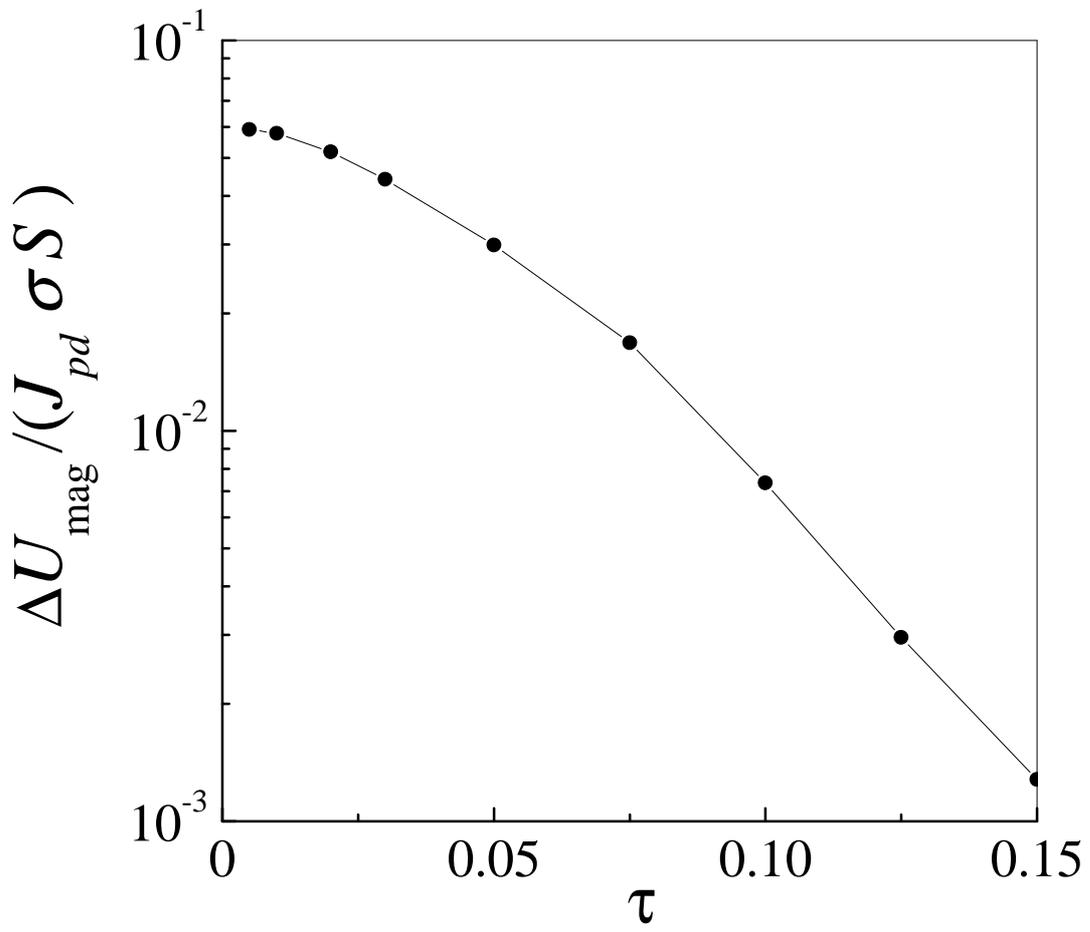
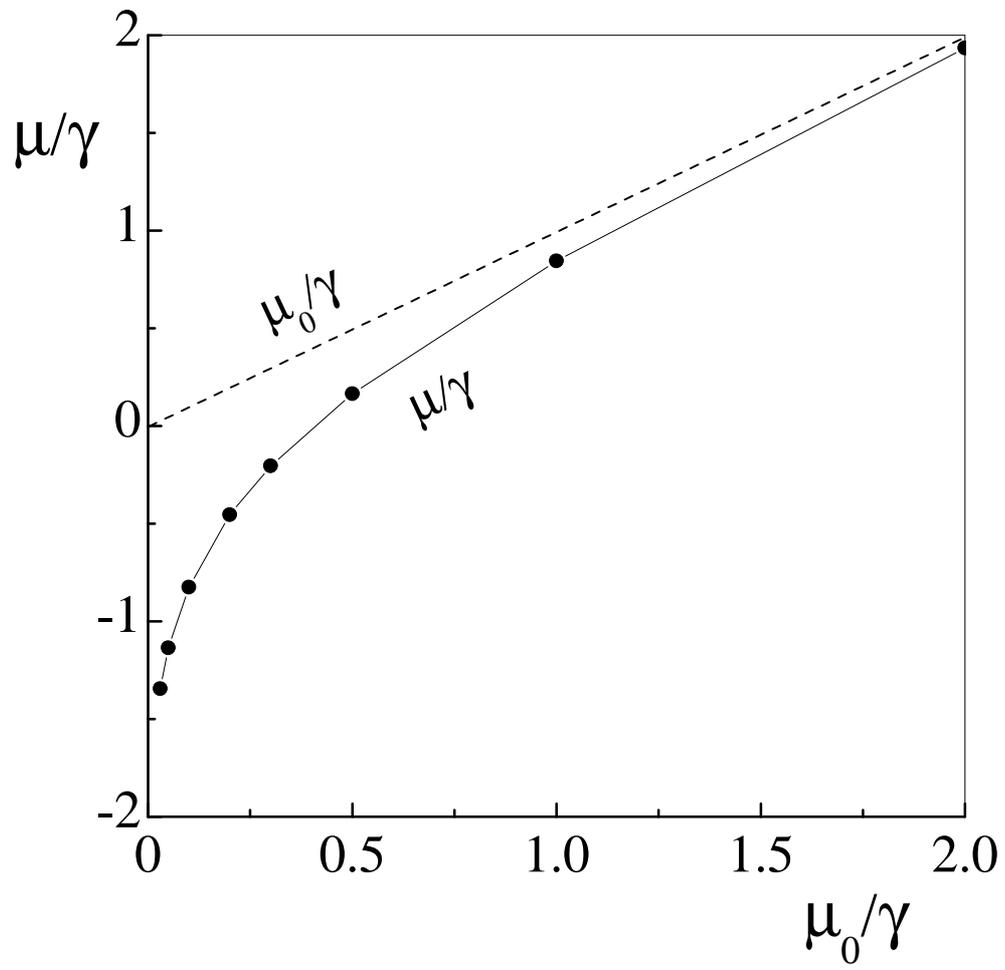
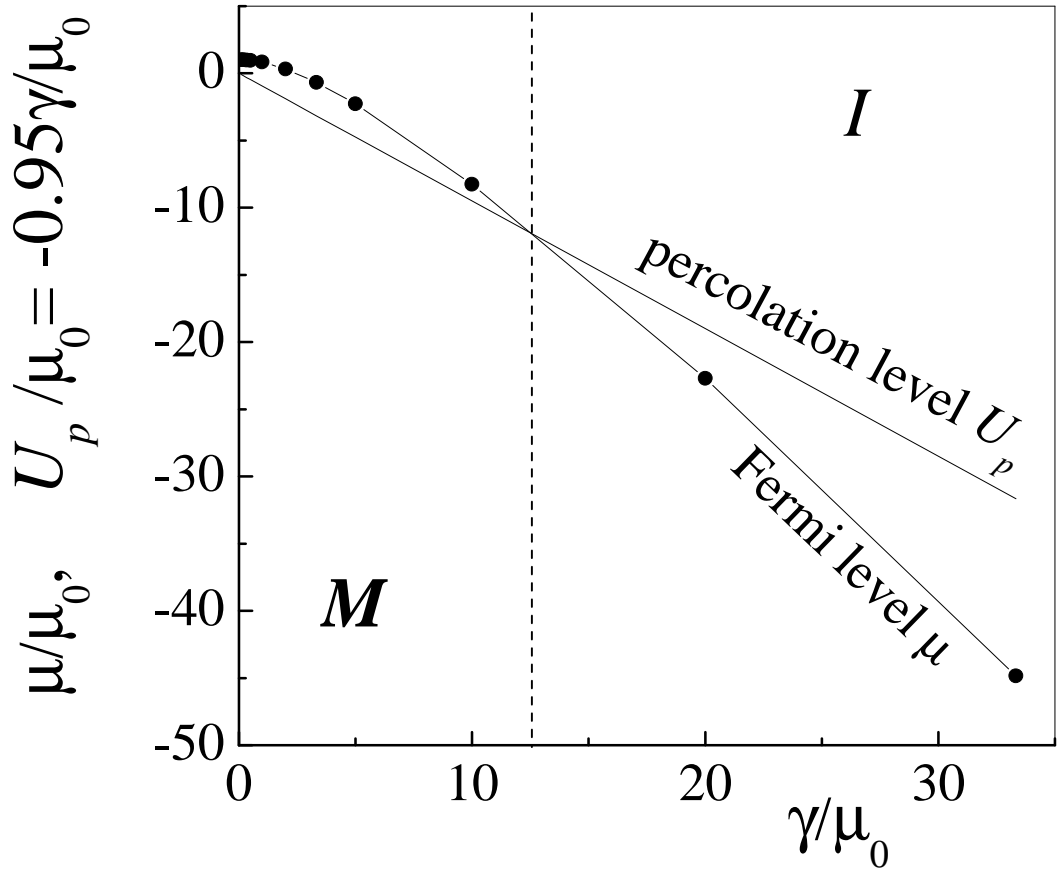
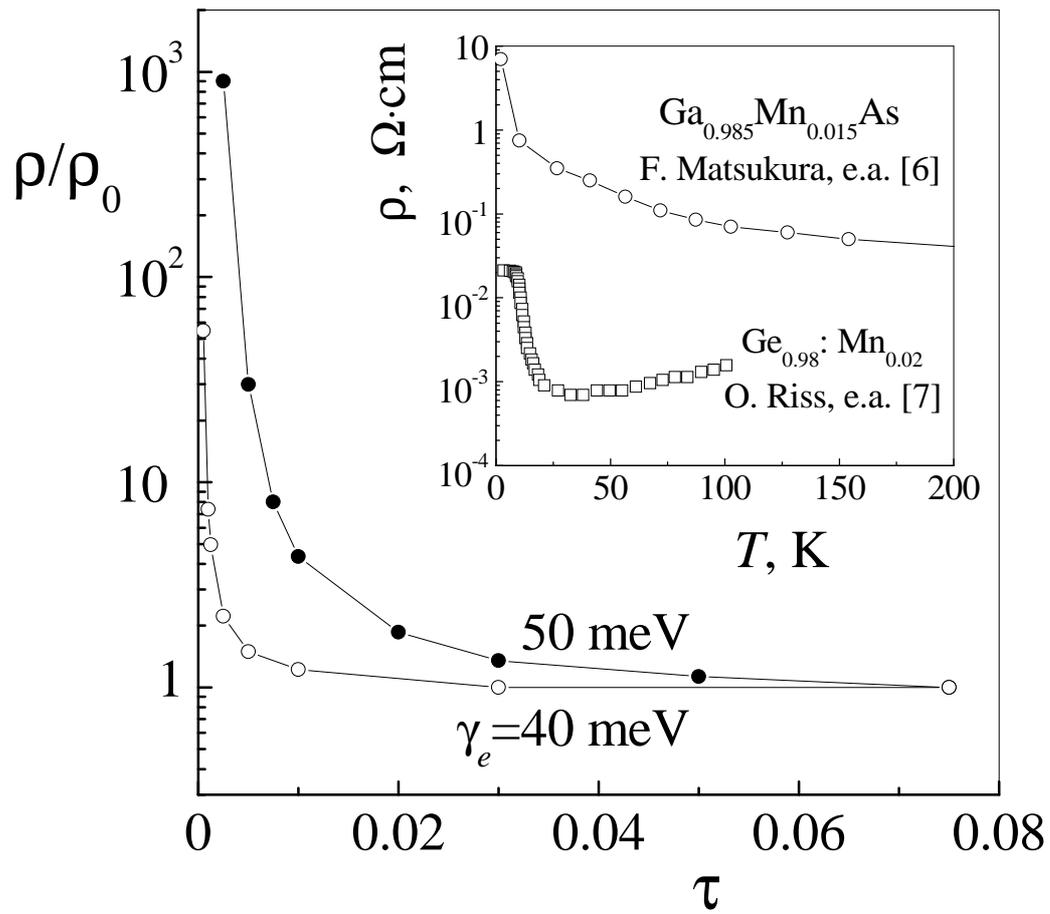
\end{document}